\documentclass{article}

\usepackage{arxiv}

\usepackage[utf8]{inputenc} 
\usepackage[T1]{fontenc}    
\usepackage[hidelinks]{hyperref}       
\usepackage{url}            
\usepackage{booktabs}       
\usepackage{amsfonts}       
\usepackage{nicefrac}       
\usepackage{microtype}      
\usepackage{graphicx}
\usepackage{natbib}
\usepackage{doi}

\RequirePackage{inputenc,crop,graphicx,amsmath,array,color,amssymb,flushend,stfloats,amsthm,chngpage,times,datetime,epstopdf} 
\usepackage{algorithm}
\usepackage{algpseudocode}
\usepackage{caption}
\usepackage{subcaption}
\algnewcommand{\algorithmicand}{\textbf{ and }}
\algnewcommand{\algorithmicor}{\textbf{ or }}
\algnewcommand{\OR}{\algorithmicor}
\algnewcommand{\AND}{\algorithmicand}
\algnewcommand{\var}{\texttt}
\algnewcommand\True{\textbf{true}\space}
\algnewcommand\False{\textbf{false}\space}

\title{A Bayesian dynamic stopping method for evoked response brain-computer interfacing}
\renewcommand{\headeright}{}  
\renewcommand{\undertitle}{}  

\date{} 					

\author{
    \href{https://orcid.org/0000-0002-0919-8565}{\includegraphics[scale=0.06]{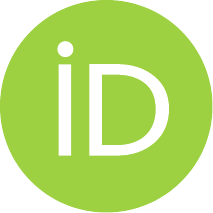}}\hspace{1mm}Sara Ahmadi\thanks{Corresponding author} \\
    Donders Institute for Brain, Cognition and Behaviour \\
    Radboud University \\
    Nijmegen, the Netherlands \\
    \texttt{sara.ahmadi@donders.ru.nl} \\
    \AND
    \hspace{1mm}Peter Desain \\
    (1) Donders Institute for Brain, Cognition and Behaviour \\
    Radboud University \\
    Nijmegen, the Netherlands \\
    (2) MindAffect \\ 
    Ede, the Netherlands \\
    \texttt{p.desain@donders.ru.nl} \\
    \AND
    \href{https://orcid.org/0000-0002-6264-0367}{\includegraphics[scale=0.06]{orcid.pdf}}\hspace{1mm}Jordy Thielen \\
    Donders Institute for Brain, Cognition and Behaviour\\
    Radboud University\\
    Nijmegen, the Netherlands \\
    \texttt{ jordy.thielen@donders.ru.nl} \\
}

\hypersetup{
    pdftitle={A Bayesian dynamic stopping method for evoked response brain-computer interfacing},
    pdfsubject={},
    pdfauthor={Sara Ahmadi, Peter Desain, Jordy Thielen},
    pdfkeywords={Bayes test, brain-computer interfacing, dynamic stopping, early stopping, visual evoked potentials},
}

\begin{document}
\maketitle

\begin{abstract}
As brain-computer interfacing (BCI) systems transition from assistive technology to more diverse applications, their speed, reliability, and user experience become increasingly important. Dynamic stopping methods enhance BCI system speed by deciding at any moment whether to output a result or wait for more information. Such approach leverages trial variance, allowing good trials to be detected earlier, thereby speeding up the process without significantly compromising accuracy. 
Existing dynamic stopping algorithms typically optimize measures such as symbols per minute (SPM) and information transfer rate (ITR). However, these metrics may not accurately reflect system performance for specific applications or user types. Moreover, many methods depend on arbitrary thresholds or parameters that require extensive training data.
We propose a model-based approach that takes advantage of the analytical knowledge that we have about the underlying classification model. By using a risk minimisation approach, our model allows precise control over the types of errors and the balance between precision and speed. This adaptability makes it ideal for customizing BCI systems to meet the diverse needs of various applications.
We validate our proposed method on a publicly available dataset, comparing it with established static and dynamic stopping methods. Our results demonstrate that our approach offers a broad range of accuracy-speed trade-offs and achieves higher precision than baseline stopping methods.
\end{abstract}


\section{Introduction}
The field of brain-computer interfacing (BCI) has witnessed remarkable advancements in recent years, offering exciting opportunities to enable novel communication pathways between the human brain and external devices. Typically, BCIs rely on electroencephalography (EEG), because it is a neuroimaging technique that can non-invasively capture electrical activity generated by the brain and it is accessible for a broad range of applications. While being practical, the adoption of EEG in BCI also introduces significant challenges, primarily stemming from an inherently low signal-to-noise ratio (SNR) associated with EEG recordings. This makes it difficult to achieve a high classification accuracy and speed in EEG-based BCI systems, leading researchers to navigate intricate terrain in neuroscience, signal processing, and machine learning to harness the full potential of this technology. 

The selection of the most relevant brain signal features in EEG, the so-called neural signatures, is a pivotal aspect of BCI design, dictating the efficacy and responsiveness of the system. Neural signatures in BCI can be broadly categorized into evoked and oscillatory signals. Evoked signals, including the event-related potential (ERP), steady-state visually evoked potential (SSVEP), and code-modulated visual evoked potential (c-VEP), represent distinct neural responses that can be reliably triggered by external stimuli and have achieved very high performances in BCIs for communication and control~\citep{gao2014}. 

An ERP is characterized by transient changes in electrical activity time-locked to specific external events. For BCI, typically the P300 ERP is used~\citep{fazel2012}. On the other hand, the SSVEP is often used, which is hypothesized to capitalize on the brain's entrainment to visual stimuli flickering at specific frequencies~\citep{tsoneva2015}, so-called frequency-tagging~\citep{vialatte2010}. Finally, the c-VEP, a relatively new but promising neural signature, involves encoding information in the pseudo-random visual stimuli to evoke specific responses, so-called noise-tagging~\citep{martinez2021}. The exploration of these evoked signals holds the key to refining the precision and versatility of more effective and adaptive BCI.

Despite the promising potential of evoked signals in BCI, the persistent challenge of a low SNR in EEG measurements remains a significant hurdle. One strategy to mitigate this limitation involves extending the duration of the visual stimulation, allowing for a more robust extraction of evoked responses from the EEG signal. Indeed, it seems that BCIs that rely on visual evoked potential (VEP) paradigms do not suffer from BCI inefficiency~\citep{volosyak2020}. However, this remedy does introduce a critical trade-off between achieving a high classification accuracy and maintaining a desirable level of speed when using a BCI. The delicate balance between trial duration and classification accuracy necessitates careful consideration in BCI design, as trials with prolonged stimulation periods may impede the real-time responsiveness required for seamless and efficient user control. 

The optimal balance between classification accuracy and speed in BCI systems is inherently dependent on the specific application at hand. As the spectrum of BCI applications continues to diversify, it becomes imperative for these interfaces to cater to a range of performance requirements tailored to distinct purposes. Consider, for instance, a BCI designed for a specific brain-controlled alarm to be raised by a patient in case of emergency. In this context, next to the overall high classification accuracy, minimizing the misses, i.e., the false negative rate, is of most importance. Conversely, in the context of a BCI developed for communication, a higher speed may take precedence, allowing for real-time interaction even at the expense of slightly lower accuracy, as they can be corrected for by the user or by post-processing the output for example using a language model \citep{gembler2019multi,gembler2019dynamic}. Thus in the scenario of a BCI speller, the inherent trade-off between accuracy and speed becomes a nuanced consideration, emphasizing the need for adaptive BCI systems that can be tailored to the specific demands of varied applications.  Only then the utility across diverse application domains and users can be maximised.

While many conventional BCI systems adhere to fixed trial lengths, allowing users to issue commands only at predetermined intervals and leveraging a constant amount of EEG data, more sophisticated methods have emerged to optimize trial duration dynamically~\citep{schreuder2013}. Within this domain, two distinct categories, namely static stopping and dynamic stopping, have garnered attention for their potential to enhance BCI performance. In static-stopping approaches, also referred to as fixed-stopping, the system learns an optimal stopping time based on training data, establishing a uniform termination point for all trials during online use. On the contrary, dynamic stopping, also known as early stopping or adaptive stimulation, introduces a (potentially calibrated) trial-by-trial decision-making process. In this paradigm, the system evaluates the level of confidence attained during each trial and determines whether the trial should be stopped and the classification be emitted. It is most useful if the set criterion and the performance reached are interpretable measures, like the expected probability of a mistake, accuracy, etc. If so, we call the method inherently calibrated, in contrast to methods that use an arbitrary score and a criterion that has to be gathered from training data. Stopping procedures introduce a valuable dimension to BCI design, offering avenues to minimize selection or stimulation duration while optimizing relevant performance metrics, thereby contributing to the continued evolution and refinement of BCI technology.

The literature on ERP BCIs has seen the proposal of various stopping methods, as extensively reviewed and compared by~\cite{schreuder2013}. One straightforward approach mandates a minimum number of consecutive iterations with identical class predictions before finalizing the classification~\citep{jin2011}. Another method accumulates classification scores over iterations per class, accepting the classification if the largest sum surpasses a learned threshold~\citep{liu2010}. In contrast, \citet{lenhardt2008} employ a dual-criteria system, requiring both the sum of classification scores to reach a learned threshold and the ratio of the best versus the second-best score to exceed a learned threshold, a multiplicative version of a margin rule. \citet{schreuder2011} propose a method that tests whether the rank difference between the best and second-best median scores is greater than a learned threshold. \citet{zhang2008} accumulate evidence for each class as the attended one, accepting the classification if the posterior probability of a class reaches a predetermined threshold. \citet{hohne2010} utilize Welch's t-test on the scores of the best versus all other classes and emit the classification when significance is reached. \citet{throckmorton2013} use a kernel density estimate to determine the probability density function for each class, accepting the classification if a predetermined threshold is reached. Finally, \citet{bianchi2019} consider the distribution of scores with respect to the separating hyperplane. 

Various stopping procedures have been proposed in the literature to optimize the termination of trials in c-VEP BCIs. One straightforward approach involves thresholding the maximum correlation ($\rho_1 \geq 0.4$) without the need for additional calibration~\citep{spuler2017}. Other non-calibrated methods fit at each potential stopping time a Beta~\citep{thielen2017, thielen2021} or Normal~\citep{martinez2023a, martinez2023b} distribution to non-maximum correlations, and threshold $\rho_1$ based on the likelihood that it is an outlier in the distribution. Conversely, calibrated methods employ logistic regression to map the correlation vector to correct or incorrect classifications, with the resulting probability thresholded to emit a classification~\citep{sato2015}. Other calibrated procedures use the margin between the best and second-best class ($\rho_1 - \rho_2$), i.e., the difference between the top two correlations~\citep{thielen2015, verbaarschot2021, gembler2019, gembler2020} or the distance to the decision hyperplane of a one-class support vector machine~\citep{riechmann2015}. Finally, two methods integrate classifications over time, one utilizing $p$-values of correlations with a learned threshold~\citep{nagel2019b}, and the other emitting a classification after a predefined number of consecutive identical classifications~\citep{castillos2023}. The wild variety of stopping procedures on the one hand contributes to the refinement of c-VEP BCIs, offering flexibility in adapting trial durations to optimize performance metrics. On the other hand, the differences make their evaluation hard and makes one desire the availability of a method that is best in all occasions.

While the aforementioned stopping methods in principle optimize BCI performance, they are severely limited. Most of them dependent on hyper-parameters that necessitate optimization with training data. Typically, this involves performing a cross-validated grid-search with candidate hyper-parameter values, and selecting those that maximize a performance metric like classification accuracy, information transfer rate (ITR), or symbols per minute (SPM). However, a potential drawback emerges from the user's perspective, as the chosen performance metric may not accurately reflect the system's efficacy for a specific application or user type. As mentioned earlier, in some applications the cost of different types of errors can be different. This means that the types of errors that the stopping method makes, regardless of its average performance, contribute to the effectiveness of the methods in terms of user-oriented efficiency. Therefore a more suitable approach is to optimize risk rather than average accuracy or average trial duration. 

Another limitation of the existing stopping methods is that they only decide based on the confidence of some similarity score and ignore possible differences between the prior probability of each output. For example in a speller system, the trials selecting frequent letters can be decided based on a much lower confidence score compared to non-frequent letters. 

To address these limitations, we propose a dynamic stopping method that focuses on minimizing the risk associated with each type of error. The model also makes it possible to incorporate different prior probabilities for target outputs. Through an analytical model-based approach, we offer a more tailored and user-centric optimization that allows one to control the system's behaviour using the algorithm's adjustable parameters. In turn, the BCI can be aligned with the nuanced requirements of different user preferences and applications.

In this paper, firstly, we show the full mathematical derivation of the proposed Bayesian dynamic stopping (BDS) method. Secondly, using a published open-access c-VEP dataset, we compare the proposed algorithm with other stopping methods and do so both in terms of conventional accuracy-time metrics as well as relevance-based performance criteria such as precision, recall and F-score. Despite that the model allows for non-equal priors, in this study, we focus on the risk optimization aspect and limit the formulation to equal prior scenarios.

\section{Material and Methods}

\subsection{Dataset}
We assessed the effectiveness of the proposed Bayesian dynamic stopping (BDS) framework using an openly accessible c-VEP dataset~\citep{thielen2015dataset}. Detailed information about this dataset can be found in the original publication~\citep{thielen2015}. For our study, we exclusively used the part of the dataset labeled as `fixed-length testing trials', where $12$ participants were involved in a copy-spelling task. EEG data were recorded from $64$ Ag/AgCl active electrodes arranged according to the international 10-10 system, amplified by a Biosemi ActiveTwo amplifier, and sampled at a frequency of $2048$\,Hz.

During the experiment, participants interacted with a $6 \times 6$ matrix speller presented on a $24$\,in BENQ XL2420T LED monitor, with a refresh rate of $120$\,Hz and a resolution of $1920 \times 1080$\,pix. The $36$ cells within the matrix were displayed against a mean-luminance gray background. Each cell underwent luminance modulation using a pseudo-random binary sequence at full contrast, where a value of 1 represented a white background and 0 indicated a black background.

The stimulus sequences were carefully optimized subsets derived from a collection of Gold codes~\citep{gold1967}. Specifically, using training data, a participant-specific subset was chosen to minimize the maximum correlation between template responses within the subset~\citep{thielen2015}. Furthermore, these selected codes were assigned to cells in the matrix in a manner that ensured neighboring sequences maintained minimal correlation in their responses~\citep{thielen2015}. The chosen sequences were modulated to include flashes of only two durations: a short flash lasting $8.33$\,ms and a long flash lasting $16.67$\,ms. With a sequence length of $126 = 2 \times (2^6-1)$\,bits and a presentation rate of $120$\,Hz, one cycle lasted $1.05$\,s.

Participants completed three identical runs, each consisting of $36$ trials, one for each of the $36$ cells, presented in a randomized order. Each trial started with a $1$-second cue, highlighted in green, indicating the target cell. Following this, all cells began flashing with their respective stimulus sequences for a duration of $4.2$\,s, during which participants maintained fixation on the target cell. Subsequently, participants received feedback from an online classifier, with the predicted cell being highlighted in green for $1$\,s.

In summary, for each participant, the dataset contained $108$ trials of a $4.2$-second trial duration, including 3 repetitions of each of the $36$ stimuli.

\subsection{Bayesian Dynamic Stopping}\label{sec:Bayesian Early Stopping}
In this section, we introduce the Bayesian dynamic stopping (BDS) framework. Let us consider $\mathbf{x} \in \mathbb{R}^{T}$ as the EEG data for a single trial and a single (e.g., spatially filtered, virtual) channel, comprising $T$ samples. We conceptualize this data as the composite of the assumed underlying template EEG response $\mathbf{t}_{y} \in \mathbb{R}^{T}$, representing the source signal for the attended stimulus with a target label $y$, and a component of noise $\boldsymbol{\epsilon} \in \mathbb{R}^T$ drawn from a normal distribution $\mathcal{N}(0, \sigma)$, where $\sigma \in \mathbb{R}$ denotes the standard deviation of the noise. Here, $\alpha \in \mathbb{R}$ represents a scaling factor.
\begin{equation}\label{eq:forward_model}
    \mathbf{x} = \alpha\mathbf{t}_y + \boldsymbol{\epsilon}
\end{equation}

Let us define a similarity score $f_i$, calculated as the inner product between the EEG data $\mathbf{x}$, corresponding to the true class label $y$, and each of the templates $\mathbf{t}_i$, representing the (expected) source signal of the candidate stimuli $i \in \{1, \dots, N\}$ from a set of $N$ classes:
\begin{equation}\label{eq:score}
    f_i
    = \mathbf{x}^\top\mathbf{t}_i
    = (\alpha\mathbf{t}_{y} + \boldsymbol{\epsilon})^\top\mathbf{t}_i
    = \alpha\mathbf{t}_{y}^\top\mathbf{t}_i + \boldsymbol{\epsilon}^\top\mathbf{t}_i
\end{equation}

Since $\boldsymbol{\epsilon}$ is an uncorrelated Gaussian distributed noise source, $f_i$ is also Gaussian distributed:
\begin{equation}\label{eq:score_distribution}
    f_i \sim \mathcal{N}(\alpha\mathbf{t}_y^\top\mathbf{t}_i, \sigma\|\mathbf{t}_i\|)
\end{equation}

Using the assumption that the observed EEG data $\mathbf{x}$ originated from class $y$, we can deduce that the similarity score $f_i$ follows a distribution that is either a target or a non-target Gaussian:
\begin{equation}\label{eq:target_non_target_score_distribution}
    f_i \sim
    \begin{cases}
        \mathcal{N}(\alpha b_1, \sigma_1) & \text{if } i = y \\
        \mathcal{N}(\alpha b_0, \sigma_0) & \text{otherwise} \\
    \end{cases}
\end{equation}
Here, the target distribution has a mean of $\alpha b_1 = \alpha\mathbf{t}_{y}^\top\mathbf{t}_{y}$ and a standard deviation of $\sigma_1=\sigma\|\mathbf{t}_y\|$, and the non-target distribution has a mean of $\alpha b_0 = \alpha\mathbf{t}_{y}^\top\mathbf{t}_i$ and a standard deviation of  $\sigma_0=\sigma\|\mathbf{t}_i\|$.

At this point, we encounter two challenges. Firstly, we have to identify the most probable true target. Secondly, we need to decide whether we possess sufficient certainty to emit this true target, or to acquire more data, specifically, to increase the number of samples $T$.

Let us first consider the simpler two-class problem, where $N=2$. Assuming that the target class is $y=1$ with a prior probability of $p_1$, and the non-target class is $i=0$ with a prior probability of $p_0$, we can define two hypotheses:
\begin{itemize}\label{eq:hypothesis}
    \item $\mathcal{H}_1$: the observed score $f_i$ originates from the target distribution $\mathcal{N}(\alpha b_1, \sigma_1)$.
    \item $\mathcal{H}_0$: the observed score $f_i$ stems from the non-target distribution $\mathcal{N}(\alpha b_0, \sigma_0)$.
\end{itemize}

To determine the correct hypothesis, we use the Bayes criterion, which relies on two fundamental assumptions. Firstly, we possess knowledge of the prior probabilities associated with the source outputs, denoted as $p_1$ and $p_0$. Secondly, we know the costs assigned to each course of action, specified as follows:
\begin{itemize}\label{eq:cost}
    \item $c_{00}$ is the cost of choosing $\mathcal{H}_0$ when $\mathcal{H}_0$ is true, i.e., no detection or true negative.
    \item $c_{01}$ is the cost of choosing $\mathcal{H}_0$ when $\mathcal{H}_1$ is true, i.e., false reject or false negative.
    \item $c_{10}$ is the cost of choosing $\mathcal{H}_1$ when $\mathcal{H}_0$ is true, i.e., false accept or false positive.
    \item $c_{11}$ is the cost of choosing $\mathcal{H}_1$ when $\mathcal{H}_1$ is true, i.e., detection or true positive.
\end{itemize}

The Bayes test aims to determine the decision region in a manner that minimizes the risk $\mathcal{R}$, i.e., the Bayes criterion~\citep{vantrees2004}. Denoting the expected value of the cost as the risk $\mathcal{R}$, we have:
\begin{equation}\label{eq:risk}
\begin{split}
    \mathcal{R} = 
    &c_{00}p_0 P(\text{say }\mathcal{H}_0 | \mathcal{H}_0 \text{ is true}) +\\
    &c_{01}p_1 P(\text{say }\mathcal{H}_0 | \mathcal{H}_1 \text{ is true}) +\\
    &c_{10}p_0 P(\text{say }\mathcal{H}_1 | \mathcal{H}_0 \text{ is true}) +\\
    &c_{11}p_1 P(\text{say }\mathcal{H}_1 | \mathcal{H}_1 \text{ is true})
\end{split}
\end{equation}

The Bayes test results in a likelihood ratio test, as discussed in Section\,2.2 of~\citet{vantrees2004}:
\begin{equation}\label{eq:likelihood_ratio_test}
    \Lambda(f_i)
    \begin{matrix}
        \overset{\mathcal{H}_1}{>} \\ 
        \underset{\mathcal{H}_0}{<}
    \end{matrix} \ \frac{p_0(c_{10} - c_{00})}{p_1(c_{01} - c_{11})}
\end{equation}
where the likelihood ratio $\Lambda(f_i)$ is defined as follows:
\begin{equation}\label{eq:likelihood_ratio}
    \Lambda(f_i) 
    = \frac{P(f_i | \mathbf{t}_1)}{P(f_i | \mathbf{t}_0)} 
    = \frac{
        \frac{1}{\sigma_1\sqrt{2\pi}}e^{-\frac{1}{2}(\frac{f_i-\alpha b_1}{\sigma_1})^2}}{
        \frac{1}{\sigma_0\sqrt{2\pi}}e^{-\frac{1}{2}(\frac{f_i-\alpha b_0}{\sigma_0})^2}}
\end{equation}
Here, dealing with $N=2$ classes and following Equation~\ref{eq:target_non_target_score_distribution}, we have $b_0=\mathbf{t}_0^\top\mathbf{t}_1$, $b_1=\mathbf{t}_1^\top\mathbf{t}_1$, $\sigma_0=\sigma\|\mathbf{t}_0\|$, and $\sigma_1=\sigma\|\mathbf{t}_1\|$.

In the context of our BCI problem setting, the risk $\mathcal{R}$ comprises only two components: the cost of a false positive (i.e., a misclassification) and the cost of a false negative (i.e., missing the target). Therefore, by setting $c_{00} = c_{11} = 0$, indicating no cost for correct detection, and introducing the cost ratio $\zeta=\frac{c_{10}}{c_{01}}$, the Bayes test can be formulated as follows:
\begin{equation}\label{eq:bayes_test}
    \ln \Lambda(f_i)
    \begin{matrix}
        \overset{\mathcal{H}_1}{>} \\ 
        \underset{\mathcal{H}_0}{<}
    \end{matrix} \ \ln \left( \frac{p_0}{p_1}\zeta \right) 
\end{equation}

Let us now transition to the more common scenario involving more than two classes. Interestingly, we can approach a multi-class problem with $N>2$ as a two-class problem by treating it as a ``one versus rest'' problem. Here, we assume that any target class conforms to the target Gaussian distribution, while all non-target classes adhere to the non-target Gaussian distribution. Consequently, we are still confronted with the same two distributions, target and non-target, albeit with their parameters estimated as the averages of their respective distributions. Specifically, assuming an equal prior for all the $N$ classes, the priors can be merged in two non-equal classes having a ratio of $N$, with the probability of a target being $p_1=\frac{1}{N}$ and the probability of a non-target being $p_0=\frac{N-1}{N}$. In this multi-class scenario, the mean and variances of the target and non-target distributions from Equation~\ref{eq:target_non_target_score_distribution} become (see Figure~\ref{fig:DistvsTime}):
\begin{align}
    \alpha b_1 &= \alpha \frac{1}{N} \sum_i^N \| \mathbf{t}_i \|^2 \label{eq:b1}\\
    \alpha b_0 &= \alpha \frac{1}{N^2 - N} \sum_i^N \sum_{j \neq i}^N \mathbf{t}_i^\top\mathbf{t}_j \label{eq:b0}\\
    \sigma_1^2 &= \sigma^2 b_1 + \frac{1}{N} \sum_i^N (\alpha \mathbf{t}_i^\top\mathbf{t}_i - \alpha b_1)^2 \label{eq:s1}\\
    \sigma_0^2 &= \sigma^2 b_1 + \frac{1}{N^2 - N} \sum_i^N \sum_{j \neq i}^N (\alpha \mathbf{t}_i^\top\mathbf{t}_j - \alpha b_0)^2 \label{eq:s0}
\end{align}

Given Equation~\ref{eq:likelihood_ratio}, we now find:
\begin{equation}\label{eq:likelihood_ratio_filled_out}
    \ln \Lambda (f_i) = 
    \ln \frac{\sigma_0}{\sigma_1} + 
    \frac{1}{2\sigma_0^2\sigma_1^2} \left( 
    (\sigma_0^2-\sigma_1^2) f_i^2 - 
    2\alpha(\sigma_1^2b_0-\sigma_0^2b_1)f_i -
    \alpha^2(\sigma_0^2b_1^2-\sigma_1^2b_0^2) \right)
\end{equation}
Now, the likelihood ratio test from Equation~\ref{eq:bayes_test} can be reformulated to the following using $a$, $b$, and $c$ as the coefficients of the quadratic formula:
\begin{equation}\label{eq:bayes_test_abc}
    a f_i^2 + b f_i + c 
    \begin{matrix}
        \overset{\mathcal{H}_1}{>} \\ 
        \underset{\mathcal{H}_0}{<}
    \end{matrix} \ 0
\end{equation}
with
\begin{align}\label{eq:abc}
    a &= \sigma_1^2 - \sigma_0^2 \\
    b &= - 2\alpha(\sigma_1^2b_0 - \sigma_0^2b_1) \\
    c &= -\alpha^2(\sigma_0^2b_1^2 + \sigma_1^2b_0^2) + 2\sigma_0^2\sigma_1^2 \ln \frac{\sigma_0}{\sigma_1(N-1)\zeta}
\end{align}

Now, the decision boundary $\eta$ is the root of the above quadratic polynomial:
\begin{equation}\label{eq:decision_boundary}
    \eta = \frac{-b + \sqrt{b^2 - 4ac}}{2a}
\end{equation}
such that the likelihood ratio test will result in the following test:
\begin{equation}\label{eq:bayes_test_eta}
    f_i \begin{matrix}
        \overset{\mathcal{H}_1}{>} \\ 
        \underset{\mathcal{H}_0}{<}
    \end{matrix} \ \eta
\end{equation}

\subsubsection{Calibrating BDS}
\begin{figure}
    \centering
    \includegraphics[width=1\textwidth]{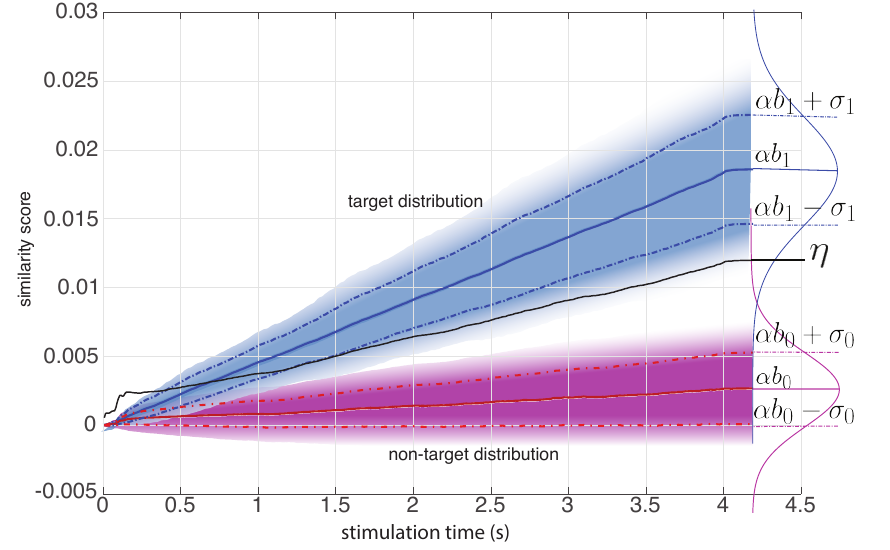}
    \caption{An example of the score distribution of target (blue) and non-target (pink) classes and how they change over stimulation time. Solid lines indicate the mean ($\alpha b_1$ and $\alpha b_0$) of the Gaussian distributions and dashed lines indicate the standard deviation ($\sigma_1$ and $\sigma_0$). The black solid line is the decision boundary $\eta$ resulting from Equation~\ref{eq:decision_boundary} with $\zeta=1$.}
    \label{fig:DistvsTime}
\end{figure}

The framework that we have presented requires calibration to estimate the distribution parameters, given a set of labeled training data. 
Firstly, the template responses $\mathbf{t}_i$ are calculated for each of the $N$ classes. For instance, for c-VEP data, one could use the `reference pipeline'~\citep{martinez2021} or `reconvolution'~\citep{thielen2015, thielen2021}.
Secondly, the template similarity score $\mathbf{t}_i^\top \mathbf{t}_j$ is calculated for each pair of stimuli at each stimulation time point, that is, for each of $S$ decision window lengths. 
Thirdly, the scaling parameter $\alpha$ is obtained from a least square error minimization between $\mathbf{x}$ and $\mathbf{t}$. If the calibration data contain multiple trials, these are concatenated as if they were one long trial.
Fourthly, a Gaussian distribution is fit on the residuals of the least squares fit, and the standard deviation of the noise $\sigma$ is estimated as the standard deviation of this residual distribution. 
Fifthly, the distribution parameters $\mathbf{b}_0\in\mathbb{R}^S$, $\mathbf{b}_1\in\mathbb{R}^S$, $\boldsymbol{\sigma}_0\in\mathbb{R}^S$ and $\boldsymbol{\sigma}_1\in\mathbb{R}^S$ are calculated according to Equation~\ref{eq:b1}--\ref{eq:s0},
and the decision boundaries $\boldsymbol\eta\in\mathbb{R}^S$ are computed following Equation~\ref{eq:abc}--\ref{eq:decision_boundary}. For an illustration of these parameters, see Figure~\ref{fig:DistvsTime}.

The decision boundary always defines the minimum risk decision area. The behaviour of BDS is controlled by the hyper-parameter cost ratio $\zeta$. With equal prior for target and non-target distributions and a cost ratio of $\zeta=1$, the decision boundary would always be at the intersection of the target and non-target distributions. Instead, as can be concluded from Equation~\ref{eq:likelihood_ratio}--\ref{eq:decision_boundary}, increasing the number of competing classes and as a result increasing the prior probability of the non-target class, results in shifting the decision boundary towards the target distribution.  

According to Equation~\ref{eq:bayes_test}, the cost ratio $\zeta$ controls the cost of the two types of error. Therefore, $\zeta=1$ means that both types of error have equal costs, hence the decision boundary will be the intersection of the two distributions. However, depending on the BCI application, the cost of false positive might be different than the cost of false negative. Imposing more cost on a false positive will result in a more reliable but slower system, while making a false negative more costly makes the system faster but more error-prone. The cost ratio parameter $\zeta$ in our formulation gives the user of the system the possibility of controlling this balance. Therefore, a cost ratio of $\zeta>1$ means the cost of false positive is higher than false negative and therefore the decision boundary moves to the right and vice versa. 

\subsubsection{Applying BDS}
Once all the model parameters are calculated, BDS can be used for dynamic stopping. At each stimulation time point, all $f_i$ for $i \in \{1, \dots, N\}$ will be compared with the decision boundary $\eta$. The index of the score that passes the test shows the winning class. If more than one score is above the threshold, the class with the highest likelihood will be chosen. If no score passes the test, the system waits for more data and repeats the test at the next stimulation time point. 

BDS makes a decision as soon as the minimum risk criterion is met. Specifically, the algorithm terminates at the earliest stimulation time point $t$ where at least one class's score $f_i$ surpasses the minimum risk decision boundary $\eta$. Furthermore, it is possible to establish a maximum trial length $t^*$, after which the most probable class label is emitted (or potentially the trial is ignored and redone). The BDS procedure is outlined in Algorithm~\ref{alg:BDS0}.

\begin{algorithm}
    \caption{BDS}\label{alg:BDS0}
    \begin{algorithmic}[1]
        \State $\var{emit} \gets \False{}$
        \State $t \gets 1$
        \While {$\var{emit} = \False$}
            \State Calculate the scores $\mathbf{f}_i(t)$ 
            \If {$\exists i: \mathbf{f}_i(t)>\eta(t)$ \OR $t = t^*$}
                \State $\hat{y} \gets \underset{i}{\arg\max}(P(\mathbf{t}_i|\mathbf{f}_y))$ 
                \State $\var{emit} \gets \True$
            \Else 
                \State $t \gets t+1$
            \EndIf
        \EndWhile
    \end{algorithmic}
\end{algorithm}

In summary, at each stimulation time point, the dynamic stopping algorithm makes a decision. If the score is below the threshold, it is a negative decision and in case the score is above the threshold, it makes a positive decision. Hence, there are four possible outcomes corresponding to the four courses of action listed in Section~\ref{sec:Bayesian Early Stopping}:
\begin{itemize}
    \item No winning class is detected while the score of the true class is not winning (true negative). 
    \item No winning class is detected while the score of the true class is winning (false negative or miss). 
    \item The winning class is detected while the score of the true class is not winning (false positive). 
    \item The winning class is detected while the score of the true class is winning (true positive or hit).
\end{itemize}
   
\subsection{Baseline Early Stopping Methods}\label{sec:Baselines}
We compared BDS with several other early stopping methods, including static and dynamic stopping techniques. Firstly, we contrasted BDS with the approach of conducting no early stopping, but instead employing a fixed trial length. We evaluated this by estimating a decoding curve for all trials and participants. The trial length could be regarded as a hyper-parameter for this method, and we assessed the accuracy achievable with each length. The trial length of 4.2\,s, which uses the full trial duration, exploits all available data in the dataset, theoretically offering an upper bound on classification accuracy.

Secondly, we contrasted BDS with three static stopping methods. Each of these methods estimated a decoding curve on the training data and optimized a specific criterion to determine an optimized stopping time for all trials within a given participant. Using a 5-fold cross-validation approach on the training data, a classifier was calibrated on the training split and then tested on all validation trials ranging from 100\,ms to 4.2\,s in 100\,ms increments.

The first static stopping method selected the stopping time at which the averaged decoding curve across folds achieved the highest accuracy for the first time. The second method chose the first instance where a predefined targeted accuracy was obtained. This targeted accuracy served as a hyper-parameter of the method. Lastly, the third method identified the stopping time at which the decoding curve achieved the highest information transfer rate (ITR).

Thirdly, we contrasted BDS with two dynamic stopping methods from previous studies: one relying on margins of classification scores~\citep{thielen2015} and the other on a Beta distribution of the classification scores~\citep{thielen2021}.

Firstly, the margin method was a calibrated approach that considered the margin between the maximum and runner-up classification scores. During calibration, it learned a margin threshold for stimulation time points ranging from 100\,ms to 4.2\,s in 100\,ms increments. These thresholds were learned to achieve a predefined targeted accuracy. During testing, the symbol associated with the maximum classification score was released as soon as the margin threshold was reached~\citep{thielen2015}.

Secondly, the Beta method did not require calibration. During testing, it estimated a Beta distribution based on all classification scores except the maximum one. Subsequently, it evaluated the probability that the maximum classification score did not originate from the estimated Beta distribution. If this likelihood surpassed a predefined targeted accuracy threshold, the classification was released~\citep{thielen2021}.

\subsection{Analysis}\label{sec:analysis}

\subsubsection{Classification}
For classification, we employ the `reconvolution CCA' template matching classifier~\citep{thielen2015, thielen2021}. Consider we have multi-channel, single-trial EEG data $\mathbf{X} \in \mathbb{R}^{C \times T}$, with $C$ channels and $T$ samples. To predict the label $\hat{y}$ for this trial, we perform template matching as follows:
\begin{equation}
    \hat{y} = \underset{i}{\arg\max}~f (\mathbf{w}^{\top}\mathbf{X}, \mathbf{r}^{\top}\mathbf{M}_i)
\end{equation}
Here, $\mathbf{w} \in \mathbb{R}^C$ is a spatial filter, $\mathbf{r} \in \mathbb{R}^M$ is a temporal response vector with $M$ samples, and $\mathbf{M}_i \in \mathbb{R}^{M \times T}$ is a structure matrix representing the onset, duration, and overlap of each event (e.g., flashes) in the $i$th stimulus sequence, where $i \in \{1, \dots, N\}$ and $N$ is the number of classes (i.e., the number of symbols in a matrix speller). 

The function $f$ represents the scoring function that calculates the similarity between the spatially filtered EEG data $\mathbf{w}^\top\mathbf{X}$ and the predicted template response for the $i$th stimulus $\mathbf{r}^\top\mathbf{M}_i$. Typically, $f$ is the Pearson's correlation coefficient~\citep{thielen2015, thielen2021}. However, in this work, we use the inner product, as BDS relies on the inner product as the similarity score.

The reconvolution CCA method requires the spatial filter $\mathbf{w}$ and temporal response $\mathbf{r}$ to be learned from labeled training data. This is achieved through canonical correlation analysis (CCA) as follows:
\begin{equation}
    \underset{\mathbf{w}, \mathbf{r}}{\arg\max}~\rho (\mathbf{w}^{\top}\mathbf{S}, \mathbf{r}^{\top}\mathbf{D})     
\end{equation}
In this formulation, $\mathbf{S} \in \mathbb{R}^{C \times KT}$ is a matrix consisting of $K$ concatenated training trials, and $\mathbf{D} \in \mathbb{R}^{M \times KT}$ is a matrix of stacked structure matrices that correspond to the labels of the training trials. CCA then identifies the spatial filter $\mathbf{w}$ and temporal response $\mathbf{r}$ that maximize the Pearson's correlation coefficient $\rho$ between the spatially filtered EEG data and the predicted template responses.

\subsubsection{Evaluation}
We evaluated the proposed BDS method using all 108 trials from each of the 12 participants in the dataset, using 5-fold cross-validation. For each fold, 4/5 of the data served as the training set, on which the classifier was calibrated, resulting in the spatial and temporal filters necessary for classification. The parameters of BDS, including the scaling factor $\alpha$, the noise standard deviation $\sigma$, the mean $b_1$, $b_0$ and standard deviation $\sigma_1$, $\sigma_0$ of the target and non-target distributions, and the decision boundary $\eta$ as a function of stimulation time, were estimated from the same training data as described in Section~\ref{sec:Bayesian Early Stopping}. The remaining 1/5 of the data was used as the test set, where trials were classified according to reconvolution CCA once the stopping decision was made by BDS.

The baseline methods were evaluated using the same 5-fold cross-validation approach. For the static methods, the optimal trial length was determined using the training set in each fold. To assess whether the choice of similarity measure influences classification performance, we evaluated both the Pearson's correlation coefficient and the inner product. Among the baseline methods, the dynamic Beta stopping method does not support the inner product as a similarity measure, because its distribution range is bounded, which is not applicable to the inner product.

For each method, we varied their respective hyper-parameters to evaluate the changes in performance. The hyper-parameter for BDS was the cost ratio $\zeta$. For the static methods, as well as the dynamic margin and Beta stopping methods, the targeted accuracy was used as the hyper-parameter.

To evaluate the performance of the early stopping methods, we considered several performance metrics. Firstly, we examined the conventional average accuracy versus average stopping time. However, since BDS aims to minimize risk rather than maximize accuracy, we also assessed performance using precision, recall, and F-score, which are relevance-based performance metrics. Precision (also known as positive predictive value) is the fraction of true positives among all the positive decisions:
\begin{equation}
    \text{precision}=\frac{\text{true positives}}{\text{true positives}+\text{false positives}}
\end{equation}
and recall (also known as sensitivity) is the ratio between the correct detections and all the detectable instances:
\begin{equation}
    \text{recall}=\frac{\text{true positives}}{\text{true positives}+\text{false negatives}}
\end{equation}
We also looked at specificity (i.e., true negative rate):
\begin{equation}
    \text{specificity}=\frac{\text{true negatives}}{\text{true negatives}+\text{false positives}}
\end{equation}

To compute the true/false positives/negatives, we examined all decisions made by the stopping procedures at each stimulation time point within each trial. With knowledge of the true label for each trial, we determined whether the highest similarity score at each stimulation time point corresponded to the true class. This enabled us to determine whether a positive or negative decision made by the stopping algorithm was true or false. Having the total number of true/false positive/negative decisions, we calculated precision, recall, specificity, and thereby the F-score:
\begin{equation}
    \text{F-score}=\frac{2 * \text{precision} * \text{recall}}{\text{precision} + \text{recall}}
\end{equation}

When interpreting these relevance-based measures, it is crucial to consider that the number of negative decisions in a stopping task far exceeds the number of positive decisions. This discrepancy arises because, for each trial, only one positive decision is made at the stopping time point, while negative decisions are made at all stimulation time points before that. 
 
\section{Results}
\begin{figure}[ht!]
    \centering
    \includegraphics[angle=90,width=0.75\textwidth]{
    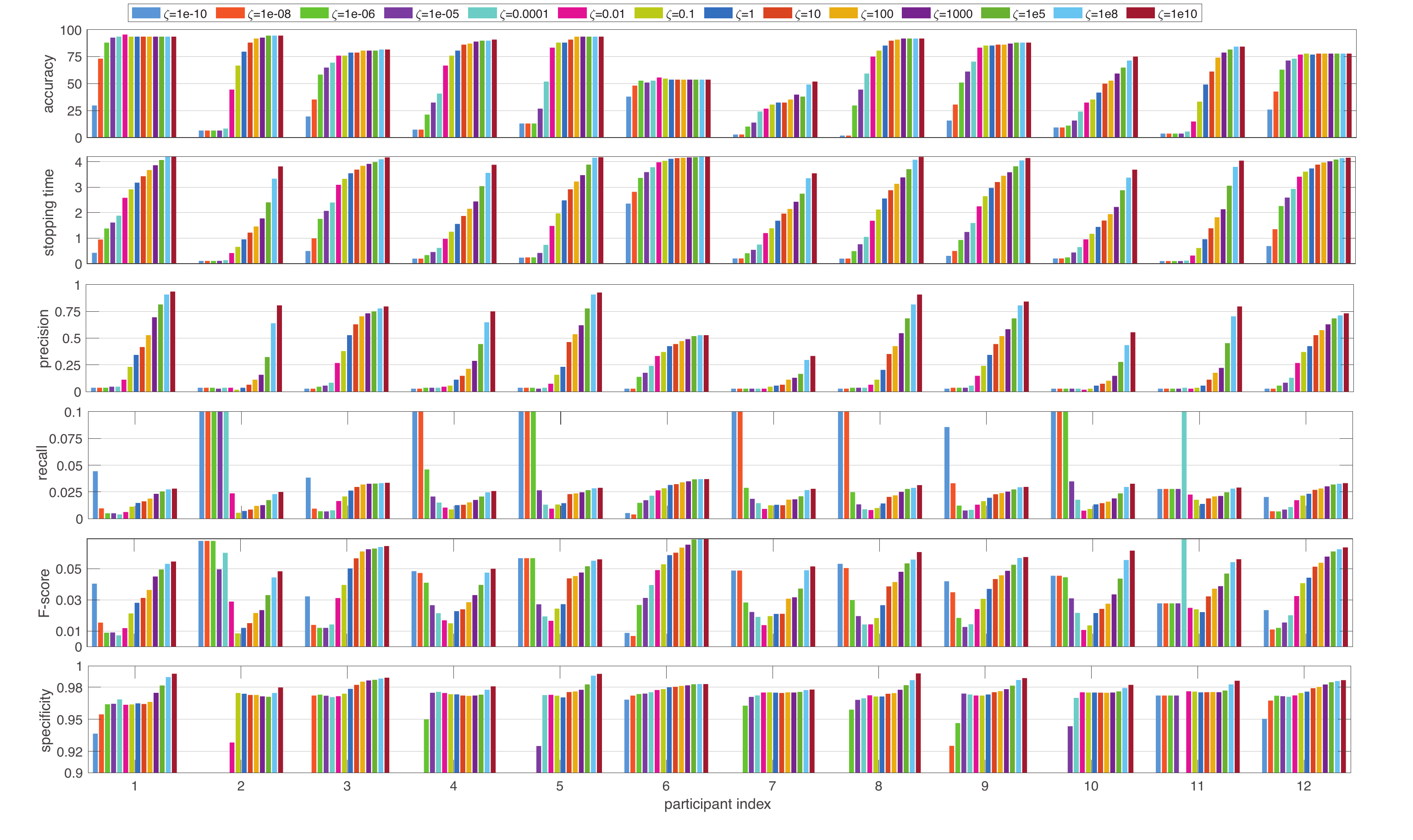}
    \caption{The performance of the Bayesian dynamic stopping averaged over the 108 trials for each participant, as the cost ratio $\zeta$ changes between 1e-10 and 1e10. The performance is measured in terms of accuracy and average stopping time as well as the relevance-based metrics precision, recall, F-score and specificity.}
    \label{fig:VaryingCR}
\end{figure}

Firstly, we investigated the performance of the proposed BDS method across various performance metrics. Figure~\ref{fig:VaryingCR} shows how the performance of BDS changes for each of the 12 participants in the dataset as the cost ratio between false positive and false negative errors $\zeta$ changes from 1e-10 to 1e10. The performance is evaluated using the six measures explained in Section~\ref{sec:analysis} and averaged over the cross-validation folds (108 trials) for each participant. 

As the cost ratio $\zeta$ increases, BDS becomes more accurate (higher accuracy) but slower (longer stopping time). The rate of these changes varies among participants. For example, for participants 1 and 6, altering the cost ratio has minimal impact on the accuracy. However, while the stopping time for participant 1 increases with a higher cost ratio, participant 6 shows much less sensitivity in stopping time. 

Investigating precision, recall, specificity and F-score reveals several trends as the cost ratio $\zeta$ increases. Precision rises more sharply than accuracy, indicating that precision is highly sensitive to the cost ratio. Recall, on the other hand, does not change monotonically. For $\zeta<1$, recall decreases as $\zeta$ increases, whereas for $\zeta>1$ an increase in $\zeta$ leads to an increase in recall. A similar pattern is observed in the F-score, suggesting that due to the larger number of negative decisions, the F-score is more influenced by recall than by precision. Finally, specificity shows smaller variations compared to other metrics, but still shows an overall increase as $\zeta$ increases. 

\begin{figure}[ht]
    \centering
    \includegraphics[width=1\textwidth]{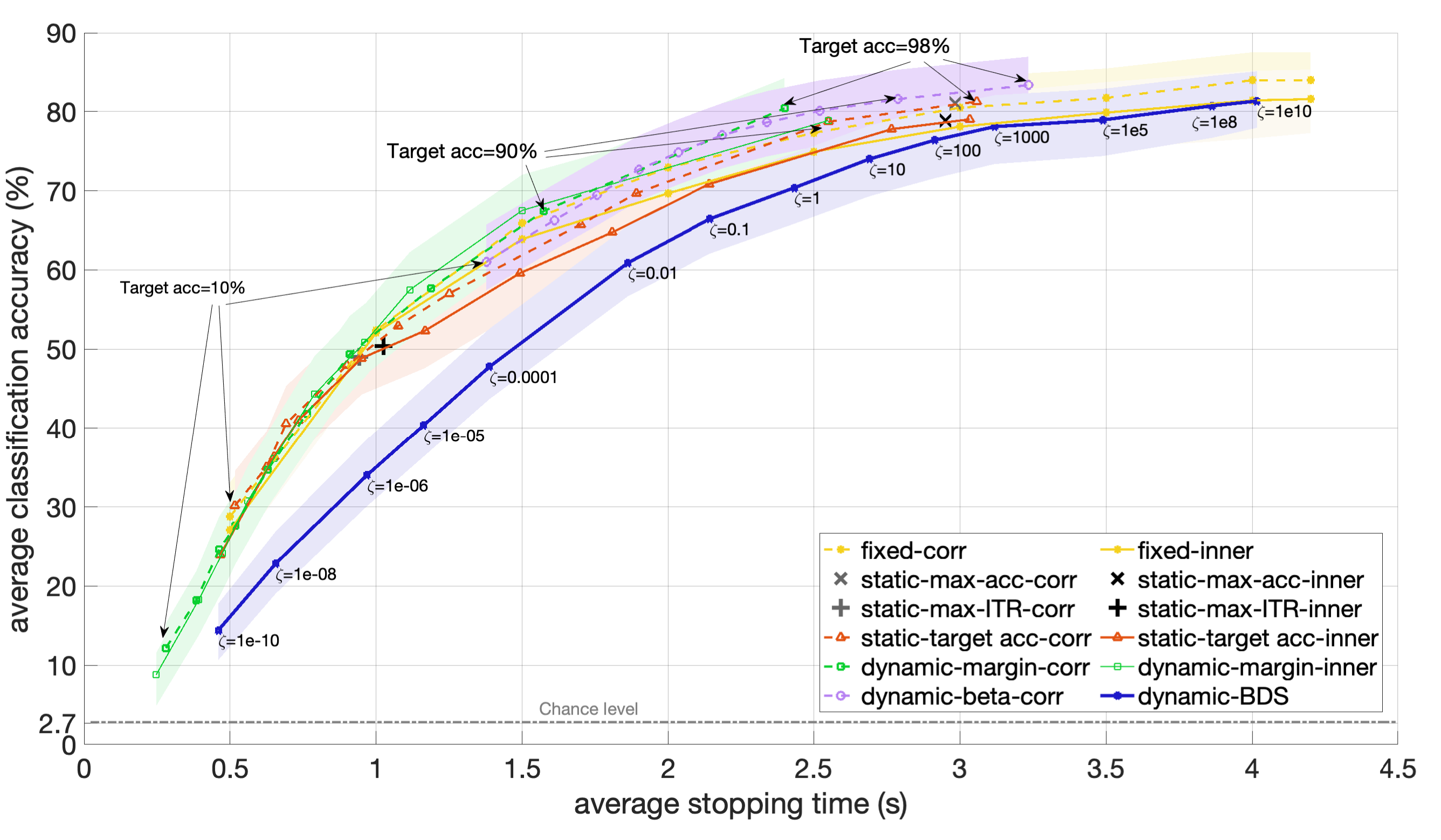}
    \caption{Accuracy versus stopping time averaged over 12 participants for BDS and static and dynamic stopping methods from the literature as the hyper-parameter of each method changes. For BDS (blue), the hyper-parameter is the cost ratio $\zeta$ ranging from $1*10^{-10}$ to $1*10^{10}$. For the static stopping methods with optimized accuracy (red), the margin (green), and beta (purple) dynamic stopping, the targeted accuracy varied between $10\%$ and $98\%$. The static methods that maximize accuracy (grey $\times$) or ITR (gray $+$) do not have a hyper-parameter and therefore appear as one point in the figure. The performance of a fixed trial length from 0.5\,s to 4.2\,s is also included for reference (yellow). All but one of the baseline methods are implemented using both correlation (dashed) and inner product (solid) as the similarity score. The shaded areas indicate the 95\% confidence interval.}
    \label{fig:AccTime-Comp}
\end{figure}

Secondly, we compared BDS with several baseline methods, including fixed, static and dynamic stopping methods, as described in Section~\ref{sec:Baselines}. Figure~\ref{fig:AccTime-Comp} presents the results in terms of average accuracy versus average stopping time, while the hyper-parameters of each method vary. 

For the fixed trial length, the hyper-parameter was the trial length, which varied between 0.5\,s and 4.2\,s. For the static stopping with accuracy optimization as well as dynamic stopping margin and beta, the hyper-parameter was the targeted accuracy, which ranged from 10\% to 90\%. The static methods that optimized accuracy or ITR did not have a hyper-parameter, hence they appear as single points in Figure~\ref{fig:AccTime-Comp} ($+$ for maximizing ITR and $x$ for maximizing accuracy). All the baseline methods, except for dynamic beta stopping, are implemented both with correlation (dashed lines) and inner product (solid lines) as similarity scores. 

As illustrated in Figure~\ref{fig:AccTime-Comp}, using the correlation as the similarity score leads to higher performance for fixed length trials and static stopping methods when the trial duration exceeds 1 second. For the dynamic margin method, the performance appears largely independent of the similarity score, except when the targeted accuracy exceeds 80\%, where the inner product results in a slightly shorter average stopping time. 

Among the dynamic stopping methods, the beta stopping method achieves the highest average accuracy in the shortest average stopping time. However, the obtained accuracy often deviates from the predefined targeted accuracy. For instance, when the targeted accuracy is set to 10\%, the beta stopping method still achieves an accuracy of 61\%, with the average stopping time not dropping below 1.3\,s. 

The dynamic margin method spans the shorter time area of the plot. With a targeted accuracy of 10\% it achieves an average stopping time of 240\,ms and an accuracy of 8.7\% accuracy (which is still higher than the theoretical chance level 2.7\%). As the targeted accuracy increases, the margin method's performance varies non-linearly. For instance, at a targeted accuracy of 98\%, the method reaches an average stopping time of 2.4\,s and an accuracy of 80.4\%. However, when the targeted accuracy is reduced to to 90\%, the resulting accuracy drops to 67\%.

The static stopping method with targeted accuracy and inner product as similarity score results in a static stopping time of 0.5\,s and an average accuracy of 30\% when the targeted accuracy is set to 10\%. For a targeted accuracy of 98\%, the obtained accuracy is 79\% with a static stopping time of 3\,s. 

In summary, each of these baseline methods covers only a limited portion of the accuracy-time plane, restricting the range within which their hyper-parameters can control them. The average accuracy values obtained by these methods for similar average stopping times are not significantly different, as indicated by the overlapping 95\% confidence intervals. Additionally, the targeted accuracy does not reliably predict the actual empirically obtained accuracy for many of the baseline methods. 

The proposed BDS tends to result in a longer average stopping time for a similar accuracy level compared to the baseline methods. Or equally, for each average stopping time (reflecting speed), BDS yields a lower accuracy level. However, the difference in accuracy between BDS and static stopping for average stopping times longer than 1.5\,s is not statistically significant, as indicated by the overlapping 95\% confidence intervals (see Figure~\ref{fig:AccTime-Comp}). Similarly, the difference in accuracy between BDS and beta stopping is not significant for average stopping times above 2.5\,s. Overall, the observed differences in accuracy were anticipated because BDS is designed to minimize risk, not error rate (hence, not to maximize accuracy). To assess the performance of each method in terms of risk, it is essential to examine precision for each method. 

\begin{figure}
    \centering
    \includegraphics[width=1\textwidth]{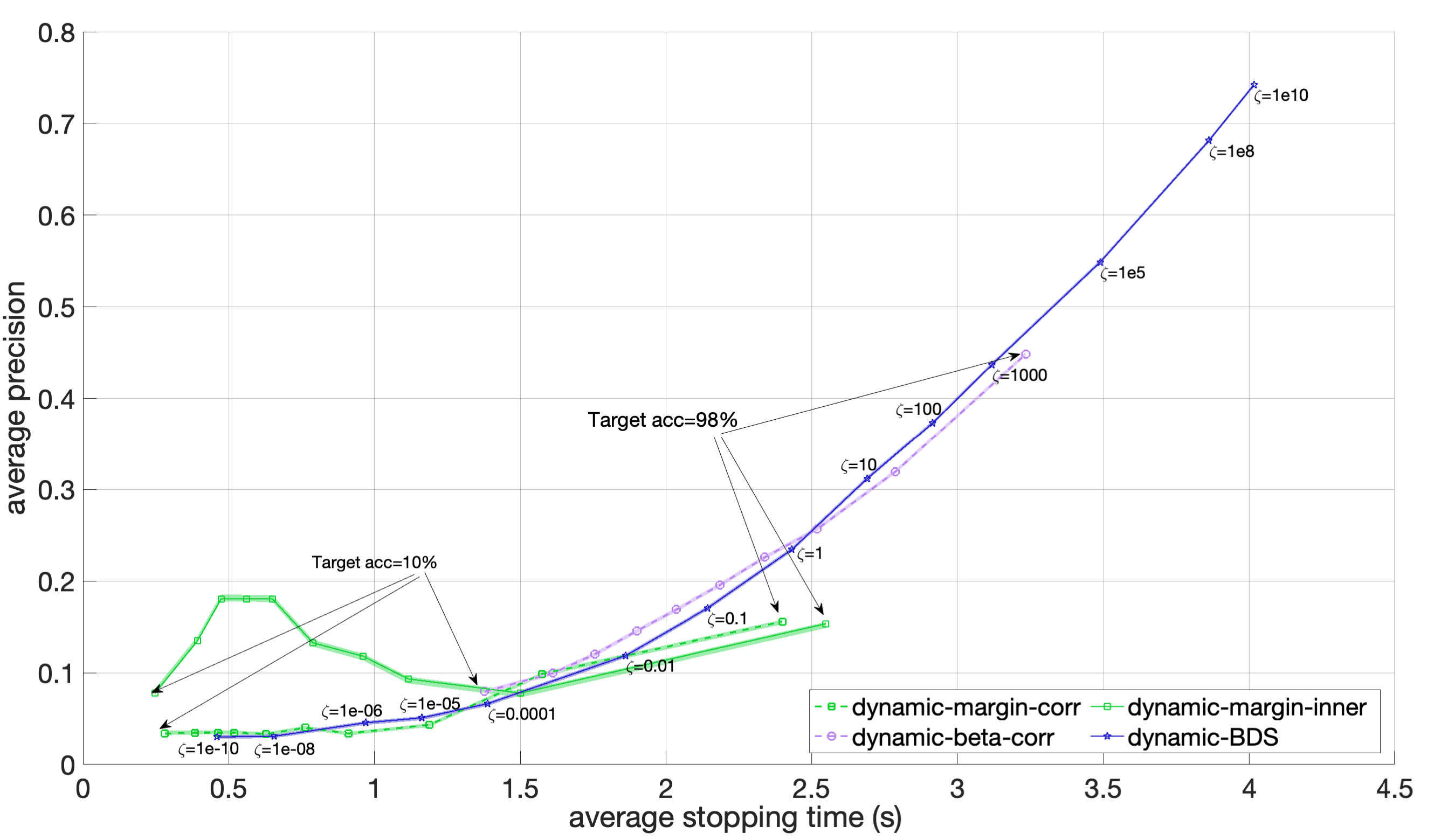}
    \caption{Precision versus stopping time averaged over 12 participants for the BDS, margin and beta dynamic stopping methods as the hyper-parameter of each method changes. For BDS (blue), the hyper-parameter is the cost ratio $\zeta$ which ranges from $1*10^{-10}$ to $1*10^{10}$. For the margin (green) and beta (purple) methods, the targeted accuracy ranges between $10\%$ and $98\%$. The margin method is implemented using both correlation (dashed) and inner product (solid) as the similarity score. The 95\% confidence interval is visualized using a shaded area, but they are in the range of 0.005 and therefore barely visible.}
    \label{fig:PrecTime-Comp}
\end{figure}

In Figure~\ref{fig:PrecTime-Comp}, the average precision across all 12 participants is plotted against the average stopping time for BDS, as well as for the two dynamic stopping baselines, i.e., the margin and beta methods. Notably, the margin method achieves a maximum precision slightly below 0.2, and the beta method obtains its highest precision of 0.44 at an accuracy of 83.3\%. Instead, BDS shows comparable precision to both the margin and beta methods for average stopping times below 1.5\,s, but beyond this, BDS surpasses the margin method in precision. Additionally, when the cost ratio $\zeta>1$, implying a higher cost for false positive than false negatives, BDS's precision begins to outperform that of the beta method as well. These observed differences are statistically significant, indicated by the very small and non-overlapping 95\% confidence intervals. This suggests that BDS excels in precision. Furthermore, BDS demonstrates the capability to increase its precision up to 0.75 by applying higher values of the cost ratio $\zeta$. 

\section{Discussion}
In this study, we proposed a model-based Bayesian dynamic stopping (BDS) method for evoked response BCIs and compared its performance with several existing static and dynamic stopping methods. The primary goal of this novel approach was to provide more control over the behaviour of the stopping method, making it suitable for various BCI applications with different requirements. For example, applications sensitive to false positives, which result in miss-classifications, need a different dynamic stopping behaviour compared to those where false negatives are more critical, which results in misses. We achieved a controllable precision-speed balance by assigning different cost values to each error type and minimizing the associated risk. 

In general, dynamic stopping methods offer various advantages over fixed trial length and static stopping methods, making them a crucial area for exploration and improvement. With dynamic stopping, we can achieve more reliable decoding by allowing more data before making decisions, unlike static methods where decisions are made at a fixed time point regardless of the trial-specific confidence level. This adaptability is particularly beneficial given the non-stationary nature of EEG data. Dynamic stopping allows for faster classification of trials with high SNR, while granting more time for those with lower SNR. Additionally, dynamic stopping ideally facilitates the incorporation of a non-control state and asynchronous control by enabling decisions only when the certainty is sufficiently high, thereby enhancing the overall system responsiveness and user experience. 

Examining the performance comparison between the baseline dynamic stopping methods and the static and fixed trial length methods (see Figure~\ref{fig:AccTime-Comp}), it is evident that dynamic stopping methods achieve accuracy levels comparable to those of static, fixed and even optimized stopping methods for a given trial length.  

While many dynamic stopping methods have been proposed for various types of evoked response BCIs, few offer inherently calibrated hyper-parameters that intuitively and effectively control the accuracy-speed balance. Additionally, many existing methods focus on optimizing metrics such as accuracy, ITR or SPM, which do not necessarily reflect user-centric performance across different applications with varying sensitivities to different error types. Our proposed method addresses this issue by providing a hyper-parameter that allows for differential treatment of error types, enabling fine-tuned control over the accuracy-speed trade-off.

We evaluated our proposed method, BDS, using an open-access c-VEP dataset including 12 participants, each with 108 trials. Performance metrics were calculated for each participant individually and then averaged across all participants. 

The per-participant results show that changing the hyper-parameter $\zeta$, which is the cost ratio between false positive and false negative errors, allows control over the accuracy and speed of BDS. However, the effect of changing the cost ratio varies among participants. Such adjustment also affects relevance-based metrics: precision, recall and specificity. When $\zeta>1$, implying a higher cost for false positive errors than for false negative errors, increasing $\zeta$ calls for a more precise classification, resulting in longer stopping times. Conversely, $\zeta<1$ implies a higher cost for false negative than for false positive errors, so decreasing $\zeta$ leads to a faster system with shorter stopping times but reduced accuracy and  precision (see Figure~\ref{fig:VaryingCR}).

Recall, by definition, is the complement of the false negative error rate. The low recall values in our results indicate an imbalance between negative and positive decisions. This is because each trial only involves one positive decision, which stops the trial, whereas negative decisions are made at all stimulation time points before the stopping time. When $\zeta>1$, increasing $\zeta$ leads to a higher number of true positives and more false negatives due to the longer waiting time. However, the increase in recall shows that the rise in true positives is larger than the increase in false negatives. Conversely, when $\zeta<1$, decreasing $\zeta$ reduces the number of true positives and false negatives. The increase in recall as $\zeta$ decreases indicates that false negatives decrease faster than true positives. This behaviour is observed in recall for most of the participants. Because of the imbalance between positive and negative decisions and the resulting disparity between the ranges of precision and recall, the F-score tends to follow the recall pattern. Finally, the high specificity values demonstrate that the number of true negatives exceeds false positives. Consequently, the effect of changing $\zeta$ is less pronounced in specificity due to this imbalance. 

We investigated the performance of BDS and the baseline methods by examining how their hyper-parameters affect accuracy and speed. An intriguing observation is that for the methods with targeted accuracy as their hyper-parameter, adjusting the targeted accuracy does change the accuracy-speed balance but does not guarantee achieving the targeted accuracy at either extreme time points. Moreover, each method spans a limited region in the accuracy-time plane, meaning that some accuracy or speed levels are unattainable for certain methods. 

In contrast, the hyper-parameter of BDS allows for a wide range, from very fast and inaccurate, to very slow and accurate classifications, controllable by the cost ratio. However, for a given speed (average stopping time), the accuracy of BDS is typically lower than the other methods. Since BDS is designed to minimize risk rather than maximize accuracy, it is more appropriate to evaluate it in terms of precision. 

Figure~\ref{fig:PrecTime-Comp} compares the dynamic methods based on precision versus stopping time. The results show that BDS achieves comparable precision to the margin and beta methods within the speed levels they share. However, interestingly, BDS can achieve a much higher precision (up to 0.75) compared to the beta method's maximum achievable precision (0.32) by assigning a higher cost to false positive errors through a larger cost ratio $\zeta$.

It is essential to approach the results of our study on the novel proposed Bayesian dynamic stopping framework with caution and consider several important limitations. Firstly, in this study, we evaluated and compared BDS using a single c-VEP dataset. While BDS shows promise, its applicability to other evoked datasets such as ERP and SSVEP requires further investigation. Additionally, we believe BDS can easily be extended to other c-VEP datasets as well, generalizing beyond the specific stimulation sequences used here, because it holds no assumptions of the underlying structure in the data.

Secondly, it is worth noting that our comparison was limited to a handful of static and dynamic stopping methods, while numerous others exist in the literature. Further research is necessary to comprehensively compare stopping methods, highlighting their underlying assumptions and application scenarios. This broader exploration can provide deeper insights into the strengths and limitations of different approaches, helping researchers and practitioners make informed decisions when selecting the most appropriate stopping method for their specific BCI applications.

Thirdly, BDS relies on template responses in the classifier, which are used to estimate the target and non-target distributions. While most standard decoding approaches involve some form of template (e.g., an event-related potential), others, such as deep learning frameworks, may not explicitly contain these templates.

 BDS needs to be calibrated using training data as it requires knowledge of the parameters for the target and non-target distributions. However, if all non-target classes can be assigned to a single non-target distribution, the required training data can be limited to data from one class only. This is especially true when using the reconvolution CCA method, which can predict template responses to unseen stimulation sequences. Not only does this mean that BDS requires only limited training data, but it also means that BDS can handle any number of classes. Investigating the sensitivity of BDS performance to the amount of training data and its empirical performance with a small number of classes remains to be addressed in future work.

Moreover, if a zero-training approach for BDS is desired, its parameters could potentially be learned on the fly as data comes in, due to its minimal requirements on the shape of the training dataset. Nevertheless, an open empirical question remains regarding how much data is needed for BDS to learn a robust and accurate dynamic stopping rule.

While BDS allows for spanning a wide range of accuracy/precision-speed balances, the current implementation does not permit setting a specific targeted performance. However, the model does offer the potential to predict error rates using the available distributions, thereby influencing the decision of the stopping algorithm. This can help achieve desired performance levels in terms of expected error rates (total or specific error types) or precision. Formalizing and implementing such an extension to this method is an avenue for future work.

Another potential advantage of using a model-based dynamic stopping method like BDS is its ability to incorporate prior probabilities for different classes. Although in this study we focused on the case of equal priors, the model formulation allows for using non-equal priors as well. This is especially valuable in applications where some classes are much more frequent than others, enabling decisions on high-probability classes with less confidence compared to low-probability classes. An example of such an application is a speller used for typing. There are studies that use a language model to post-process the output of a BCI speller, thereby increasing the typing accuracy and speed \citep{gembler2019,gembler2019dynamic}. Our proposed model offers the potential to directly integrate the language model within the dynamic stopping method, instead of using it for error correction. Integrating such priors into the stopping method is another avenue for future work.

In conclusion, BDS provides better control over false positive and false negative errors, enabling a broader range of accuracy-speed trade-offs. Additionally, BDS achieves a much higher level of precision compared to the baseline dynamic stopping methods. The proposed model opens promising directions for future work to harvest its potential more effectively.

\section*{Conflict of Interest Statement}
PD is founder of MindAffect, a company that develops EEG-based diagnosis of perceptual functions. Otherwise, the authors declare that the research was conducted in the absence of any commercial or financial relationships that could be construed as a potential conflict of interest.

\section*{Author Contributions}
SA: Conceptualization, Data curation, Formal analysis, Investigation, Methodology, Software, Validation, Visualization, Writing - original draft;
PD: Conceptualization, Project administration, Supervision, Writing - Review \& editing.
JT: Data curation, Formal analysis, Investigation, Methodology, Resources, Software, Supervision, Writing - original draft;

\section*{Acknowledgments}
This project has partially received funding from the international ALS association under grant agreement \#387 entitled, `Brain Control for ALS Patients (ATC20610)' and from the Dutch ALS foundation under grant agreement number 18-SCH-391.
The authors would like to express their gratitude towards Louis ten Bosch for his guidance in the mathematical derivation of the framework, and Jason Farquhar and Mojtaba Rostami Kandroodi for their valuable advice during the design and formalisation of the method.


\section*{Data Availability Statement}
The dataset for this study, originally recorded by~\citet{thielen2015}, can be found in the \href{https://data.ru.nl/}{Radboud Data Repository}, specifically~\citet{thielen2015dataset}. An implementation of BDS can be found in \href{https://github.com/thijor/pyntbci}{PyntBCI}.

\bibliographystyle{unsrtnat}  
\bibliography{references}

\end{document}